\documentclass{aa}
\usepackage{txfonts}
\usepackage{natbib}
\usepackage{graphicx}
\usepackage[english]{babel}
\usepackage{lscape}

\begin{document}

   \title{Formation of Multiple Populations in Globular Clusters:\\ Another Possible Scenario}

   \author{A.~A.~R. Valcarce \and M. Catelan}
   \offprints{A.~A.~R. Valcarce}
   \institute{Pontificia Universidad Cat\'olica de Chile, Departamento de Astronom\'ia y Astrof\'isica,
              Av. Vicu\~na Mackena 4860, 782-0436 Macul, Santiago, Chile\\
              \email{avalcarc@astro.puc.cl; mcatelan@astro.puc.cl}
             }
   \date{Accepted for publication in Astronomy and Astrophysics (June 2011)}

 
  \abstract
   {}
   {While chemical composition spreads are now believed to be a universal characteristic of globular clusters (GCs), not all of them present multiple populations in their color-magnitude diagrams (CMDs). Here we present a new scenario for the formation of GCs, in an attempt to qualitatively explain this otherwise intriguing observational framework.  
   }
   {Our scenario divides GCs into three groups, depending on the initial mass ($M_I$) of the progenitor structure (PS), as follows. i)~Massive PSs can retain the gas ejected by massive stars, including the ejecta of core-collapse SNe. ii)~Intermediate-mass PSs can retain at least a fraction of the fast winds of massive stars, but none of the core-collapse SNe ejecta. iii)~Low-mass PSs can only retain the slow winds of intermediate-mass stars. Members of the first group would include $\omega$~Centauri (NGC~5139), M54 (NGC~6715), M22 (NGC~6656), and Terzan~5, whereas NGC~2808 (and possibly NGC~2419) would be members of the second group. The remaining GCs which only present a spread in light elements, such as O and Na, would be members of the third group. 
   }
   {According to our scenario, the different components in $\omega$~Cen should not display a sizeable spread in age. We argue that this is consistent with the available observations. We give other simple arguments in favor of our scenario, which can be described in terms of two main analytical relations: i)~Between the actual observed ratio between first and second generation stars ($R_{SG}^{FG}$) and the fraction of first generation stars that have been lost by the GC ($S_L$); and ii)~Between $S_L$ and $M_I$. We also suggest a series of future improvements and empirical tests that may help decide whether the proposed scenario properly describes the chemical evolution of GCs.
   }
   {}

   
   \keywords{
   globular clusters: general; globular clusters: individual: NGC~2808, NGC~6715, NGC~5139, NGC~6656, Terzan~5; stars: abundances; stars: formation; stars: evolution}

   \authorrunning{Valcarce, \& Catelan}
   \titlerunning{Formation of MPs in GCs - Another Scenario}
   \maketitle
%

\section{Introduction}
\label{intro}

For many years it was believed that globular clusters (GCs) follow with excellent precision the canonical zero-order approximation for the formation of simple stellar populations, where a chemically homogeneous gas cloud collapses under the action of its own potential well forming stars of different masses at the same time. However, recent observations have shown that this hypothesis is quite far from reality. 

One of the peculiarities observed is a spread in the abundance of light elements in stars of all GCs studied to date. This phenomenon is very well represented by the O-Na anticorrelation \citep[see][for a thorough review about the chemical composition of GC stars]{Gratton_etal2004}. This anticorrelation differs in detail from one GC to the next, and its extension seems to depend on the present-day GC mass, but it does not seem to depend on the cluster's metallicity \citep{Carretta_etal2010}. 

Due to the fact that massive GCs have deeper potential wells, it is commonly thought that the spread of Fe-peak elements would only be present in the most massive GCs. However, spectroscopic evidence has revealed heavy element variations in GCs which are less massive than others which do not show a spread in these elements. This is the case of M22 (NGC~6656) and NGC~1851, with masses of $\sim 5.4\times 10^5 M_\odot$, and $\sim 5.6\times 10^5 M_\odot$, respectively, which show heavy element variations, while other GCs (such as the extensive studied M3, with a mass $\sim 8.4\times 10^5 M_\odot$) do not show this peculiarity.\footnote{Here, as in the rest of this paper, we estimate the mass of each GC using $M_{V}$ from the \citet[][revision 2010]{Harris1996} catalog and a mass-to-light ratio $M/L_{V}=3 \, M_{\odot} L_{\odot}^{-1}$, as predicted by simple stellar population models \citep{Bruzual_Charlot2003}.

} 
In fact, while M22 shows an iron spread between ${\rm [Fe/H]} \approx -1.9$ and $-1.6$, as well as spreads in other elements \citep{Pilachowski_etal1982, Lehnert_Bell_Cohen1991, DaCosta_etal2009, Marino_etal2009, fmea11}, NGC~1851 harbors two groups of stars with different [Ba/Fe] ratios, but, depending on the authors, with a uniform [Fe/H] \citep{Yong_Grundahl2008,Villanova_Geisler_Piotto2010} or with a spread in metallicity of $\Delta{\rm [Fe/H]}\sim 0.2$~dex \citep[][who also associate NGC~1851 to a merger of two separate GCs]{Carretta_etal2010a,ecea11}. From a deep study of their color-magnitude diagrams (CMDs), these GCs show that the simple stellar population hypothesis is ruled out, as revealed by an observed subgiant branch (SGB) split \citep{Marino_etal2009,Milone_etal2008}.

Another GC presenting an SGB split is NGC~6388 \citep{Piotto2009}, with a mass of $1.4\times 10^6 M_\odot$. \citet{Cassisi_etal2008} have pointed out that such SGB splits could be related to an enhancement in CNO abundances, or an internal age difference of a few Gyr. However, \citet{Villanova_Geisler_Piotto2010} have shown that for NGC~1851 the C+N+O content would be constant between the two groups with different [Ba/Fe]. Moreover, \citet{DiCriscienzo_etal2010} have pointed out that such SGB splits may also be caused by variations in the helium abundance, though only at moderately high metallicity. 

For the moment, NGC~2808 ($1.4\times 10^6 M_\odot $) seems to be a fairly special case, with at most a small spread in metals (${\rm [Fe/H]}=-1.10 \pm 0.03$) but a wide O-Na anticorrelation \citep{Carretta_etal2006}. Indeed, this GC has three populations observed down to the main sequence (MS), which suggests different helium abundances, ranging from $Y=0.25$ to $Y=0.37$ \citep{Norris2004,DAntona_etal2005,Piotto_etal2005,Piotto_etal2007}. This last peculiarity was in fact predicted by \citet{DAntona_etal2002} from a study of the cluster's horizontal branch (HB) morphology, and recently tested by \citet{Dalessandro_etal2011} using high-resolution far-UV and optical images, as well as \citet{Pasquini_etal2011} using visual and near-IR spectroscopy. Another GC which may share properties similar to NGC~2808's is NGC~2419, with a present-day mass of $1.4\times 10^6 M_\odot$ (\citeauthor{DiCriscienzo_etal2011} \citeyear{DiCriscienzo_etal2011}; \citeauthor{cd11} \citeyear{cd11}; see also \citeauthor{jcea10} \citeyear{jcea10}).

From photometry, the case of $\omega$~Cen, the most massive Galactic GC ($3\times 10^6 M_\odot$), has been known for a long time \citep{Geyer1967,Cannon_Stobie1973}, although only recent studies have unveiled some of its most intriguing properties. This GC has at least three separate, well-defined red giant branches (RGBs), an extended HB morphology, and a large number of subluminous extreme HB stars \citep[e.g.,][]{DCruz_etal1996,Whitney_etal1998,Pulone_etal1998,DCruz_etal2000,Pancino_etal2003,Rey_etal2004,Freyhammer_etal2005,Cassisi_etal2009,Bellini_etal2009,Calamida_etal2009,Calamida_etal2010}. Recent {\em Hubble Space Telescope} (HST) observations have also shown a triple MS \citep[][]{Bedin_etal2004,Bellini_etal2009,Bellini_etal2010}, in addition to at least four SGBs \citep[][]{Villanova_etal2007}. From spectroscopic studies, this GC also shows a complex behavior. Based on spectroscopically derived chemical compositions of MS stars, \citet{Piotto_etal2005} have suggested that the bluest MS is highly enriched in helium ($Y=0.35$), while metallicities of SGB and RGB stars show a wide distribution, from ${\rm [Fe/H]} \sim -2$ to $\sim -0.4$ \citep[and references therein]{Hilker_etal2004,Villanova_etal2007,Johnson_etal2009,Johnson_Pilachowski2010}. These different metallicity groups present some intriguing properties, including a smaller spread in the abundances of some elements (e.g., C, N, Ca, Ti, and Ba; see Fig.~16 in Villanova et al. 2007) at the more metal-rich end, compared to metal-poor stars in the cluster. It would thus appear that metal-rich stars were formed all at the same time, from a fairly chemically homogeneous cloud. Metal-poor stars may also have formed all at the same time, but from a chemically {\em in}homogeneous cloud. Alternatively, these metal-poor stars may have formed at different times, with material processed in different ways \citep[see also][]{rgea11}.  

Another GC with properties similar to $\omega$~Cen's, albeit at a much less extreme level, is M54 ($2.3\times 10^6 M_\odot$), in the nucleus of the Sagittarius dwarf galaxy \citep[e.g.,][]{Siegel_etal2007}. The fact that M54 is the second most massive GC in the Galactic neighborhood strongly suggests that $\omega$~Cen too may be the remnant of a formerly much larger structure \citep[e.g.,][and references therein]{Dinescu2002, Altmann_etal2005, WylieDeBoer_etal2010}. Other GCs where internal dispersions in the Fe-peak elements have recently been detected, and which may thus share some similarities in their chemical evolution with $\omega$~Cen and M54, include M22 \citep[e.g.,][]{DaCosta_etal2009,Marino_etal2009,fmea11} and Terzan~5 \citep[][]{ffea09,fdea10,loea11}.

An intriguing GC is 47~Tucanae (NGC~104), which in spite of being one of the most massive Galactic GCs ($1.4 \times 10^6 M_\odot$) presents but a moderately wide SGB~-- a feature that is interpreted by \citet{DiCriscienzo_etal2010} as being due to the presence of subpopulations with slightly different helium abundances, instead of either an internal spread in age or C+N+O. Importantly, the lack of prominent subpopulations in 47~Tuc suggests that the {\em present-day} GC mass does not correlate tightly with the presence and extent of the multiple populations observed in GC CMDs, likely indicating important mass evolution in these objects \citep[e.g.,][and references therein]{Vesperini1998, Fall_Zhang2001, Bekki_Norris2006, DAntona_etal2007, Mclaughlin_Fall2008,Conroy2011}. On the other hand, \citet{Nataf_etal2011} have very recently found a radial gradient in the luminosity of the RGB bump in 47~Tuc, interpreting this as evidence of a second generation of stars forming deep within the cluster's potential well. 

In this paper we present a scenario which tries to explain qualitatively the observed CMD peculiarities, where the main parameter that defines the {\em present-day} CMD is the {\em initial} mass ($M_I$) of the progenitor structure (PS), as opposed to the GC's present-day mass ($M_{GC}$). We assume that all GCs have lost an unknown amount of mass which is not necessarily related with $M_{GC}$, but which can be estimated assuming a simple analytical model. 

Our paper is structured as follows. First, we review in $\S$\ref{previousmodels} some of the recent scenarios for the formation of GCs. This is followed in $\S$\ref{ourmodel} by an explanation of our suggested scenario. Then in $\S$\ref{Estimations} we estimate the $M_I$ value of the PS which in our scenario has given birth to the present-day GC, based on the observed ratio between second and first generation stars ($R_{FG}^{SG}$). In this section we also discuss some open problems with our scenario. Finally, in $\S$\ref{summary} we summarize our main results.


\section{Previous Scenarios}
\label{previousmodels}

For decades there have been many attempts to establish a scenario to explain the formation and evolution of GCs \citep[e.g.,][]{Matsunami_etal1959,PikelNer1976,DiFazio_Renzini1980,DiFazio1986,TenorioTagle_etal1986,Brown_Burkert_Truran1991,Murray1992,Price_Podsiadlowski1993,Richtler_Fichtner1993,Murray_Lin1993,Brown_Burkert_Truran1995,Salaris_Weiss1997,Nakasato_Nomoto2000}, but only recently has it become clear that the common assumption that they are simple stellar populations (SSP) is generally incorrect~-- and, importantly, not only in the well-known cases of $\omega$~Cen and (more recently) M54. In fact, there is increasing evidence that virtually {\em all} GCs show some level of chemical composition inhomogeneity \citep[e.g.,][]{Carretta_etal2009a,Carretta_etal2009}, which in some cases may lead to multiple sequences in the observed CMDs. In what follows we summarize some of the scenarios that have recently been proposed to explain the new empirical evidence \citep[see ][for a more detailed comparison]{Renzini2008}. 

\subsection{\citet{Decressin_etal2007b} Scenario}
\citet{Decressin_etal2007b} have presented a scenario to explain the multiple populations observed in some massive GCs which invokes pollution from fast rotating massive stars \citep[FRMS,][]{Decressin_etal2007}. The latter are stars that have reached the critical rotation velocity, and have formed an equatorial gas disk around them. Unlike the normal winds of massive stars, which have velocities ranging from hundreds to a few thousand km/s, the equatorial disk loses mass with velocities lower than 50~km/s, which can then easily be retained by the potential wells of GCs. In this scenario, FRMS ejecta are mixed with pristine gas to form second generation (SG) stars near their massive progenitors. Even though this scenario may help explain the O-Na anticorrelation observed in all GCs, the multiple MSs observed in some GCs cannot be explained without invoking {\em discrete} helium abundances, whereas this scenario might more naturally lead to a smooth spread in $Y$, as was also pointed out by \citet{Renzini2008}. 

\subsection{\citet{DErcole_etal2008} Scenario}
\label{DErcoleModel}

In \citet{DErcole_etal2008}, the formation of multiple populations is attributed to the gas ejected by intermediate-mass AGB (or ``super-AGB'') stars \citep{Dantona_etal1983, Ventura_etal2001, Ventura_etal2002, Thoul_etal2002}, with this gas starting to collect in the GC core after the last core-collapse SN explosion. In this sense, the formation of a cooling flow is studied by the authors using one-dimensional hydrodynamical simulations. Such a cooling flow collects the gas expelled by super-AGB stars that is used to form the SG stars. Using N-body simulations, the authors study the dynamical evolution of both populations, showing that if SG stars are formed in the center of the GC, most of the expelled stars will be first generation (FG) stars, which can result in a GC dominated by SG stars.  

For their ``standard'' scenario, hydrodynamical simulations show that the star formation rate increases since 10~Myr (the time of the last core-collapse SN) until 40~Myr (the time of the first SN~Ia). However, the effect of core-collapse SNe belonging to the SG, which must stop the formation of SG stars after a few Myr, is not considered. 

Note also that every SG star that does not explode as a core-collapse SN is a possible progenitor of a type Ia SN. In fact, the more massive the progenitor, the closest the remmaining white dwarf will be to the Chandrasekhar mass \citep[e.g.,][]{Salaris_etal2009,Kalirai_etal2009}. Therefore, after the core-collapse SNe period, there must also be a period of type Ia SNe, which will inhibit the formation of stars with gas ejected by more massive intermediate-mass stars. Naturally, the frequency of these events must decrease with time, instead of remaining constant. In our opinion, this poses considerable problems for the explanation of multiple populations with high helium abundances in GCs using intermediate-mass stars. 

\subsection{\citet{Conroy_Spergel2011} Scenario}

In the \citet{Conroy_Spergel2011} scenario, the FG stars are born in an intracluster medium previously enriched in metals ({\em present-day} GC metallicity), while after some Myr the remaining gas is completely expelled by core-collapse SNe. The GC is then ready to start the accretion of the intermediate-mass AGB ejecta for several $\times 10^8$~yr. During this time, the GC is also accreting pristine material from the ambient interstellar medium (ISM), which is mixed with the AGB ejecta to form SG stars. Finally, SG core-collapse SNe, and later FG type Ia SNe, begin to explode, keeping the GC gas-free, which stops the star formation process.

Even though the accretion of material from the ISM is plausible, it is unlikely that the chemical composition of this accreted material is similar to the GC itself. Moreover, if core-collapse SN explosions cleaned the intracluster medium, it must also clean the surrounding ISM or/and increase its metallicity, implying that SG stars will be formed with different metal abundances. 

To close, we note that \citet{DErcole_etal2011} have very recently also argued against this scenario, pointing out in particular that it is not possible to form very O-poor stars through the process envisaged by \citet{Conroy_Spergel2011}. \citeauthor{DErcole_etal2011} also point out, following \citet{Renzini2008}, that the \citeauthor{Conroy_Spergel2011} scenario also fails to account for the high amount of He that is needed to explain the highly He-enriched populations that appear to be present in some GCs (see \S\ref{intro}).

\subsection{\citet{Carretta_etal2010} Scenario}
\label{carrettamodel}

\citet[and references therein]{Carretta_etal2009} have made a remarkable effort to obtain a homogeneous spectroscopic database for 17 GCs, where they have focused mainly on the O-Na anticorrelation. These data were used to divide the stellar populations in GCs into three components: i)~The primordial population, with stars with O and Na abundances similar to field stars (i.e., stars with ${\rm [Na/Fe]} < {\rm [Na/Fe]}_{\rm min} + 0.3$, where ${\rm [Na/Fe]}_{\rm min}$ is the lowest Na abundance detected in that GC); ii)~The intermediate population, with stars with ${\rm [O/Na]} > -0.9$ that do not belong to the primordial population, amounting to 60\% to 70\% of the observed stars; and iii)~The extreme population, with stars with ${\rm [O/Na]} < -0.9$, which are not presented in all GCs. 

Using these data and other global parameters characterizing the GCs in their sample, \citet{Carretta_etal2010} have proposed a scenario where GCs were formed in three different stages. First, a precursor population of stars was formed when the unborn GC (with a size of $\sim 100$~pc, and made of gas and dark matter) interacts strongly with other structures. Core-collapse SNe of this population enrich the remaining gas, and trigger the formation of the primordial population. Then, the gas ejected by primordial FRMS or super-AGB stars give rise to a gas cloud chemically enriched in the center, where the second generation (SG) of stars is born. Finally, SG core-collapse SNe clean the remaining gas, thus halting star formation. During this time, the structure has lost all its dark matter content, almost all the precursor stars, and a large fraction of the primordial population. 

While this appears to provide a promising framework, there are some points which deserve a deeper inspection. Primordial low-mass stars, which are not observed in the actual GC because they were expelled from the initial structure, must still be present in some place. Therefore, according to this scenario, a large number of metal-poor stars must be present in the field. However, to date there are only 174 known stars with ${\rm [Fe/H]} \le -3$, and 659 stars with ${\rm [Fe/H]} \le -2$ \citep[SAGA database,\footnote{\tt http://saga.sci.hokudai.ac.jp}][]{Suda_etal2008}. Alternatively, the initial mass function (IMF) must be top-heavy at low metallicities \citep{Skillman2008,Komiya_etal2010}. Note, in addition, that if the gas ejected by the precursor core-collapse SNe is retained in the initial structure, the ejecta of the primordial core-collapse SNe must also be retained, which would increase the metallicity of the SG stars unless the initial structure had already lost a large fraction of its mass in a short period of time.

\begin{figure*}[!t]
   \centering
      \includegraphics[width=18cm,height=9cm]{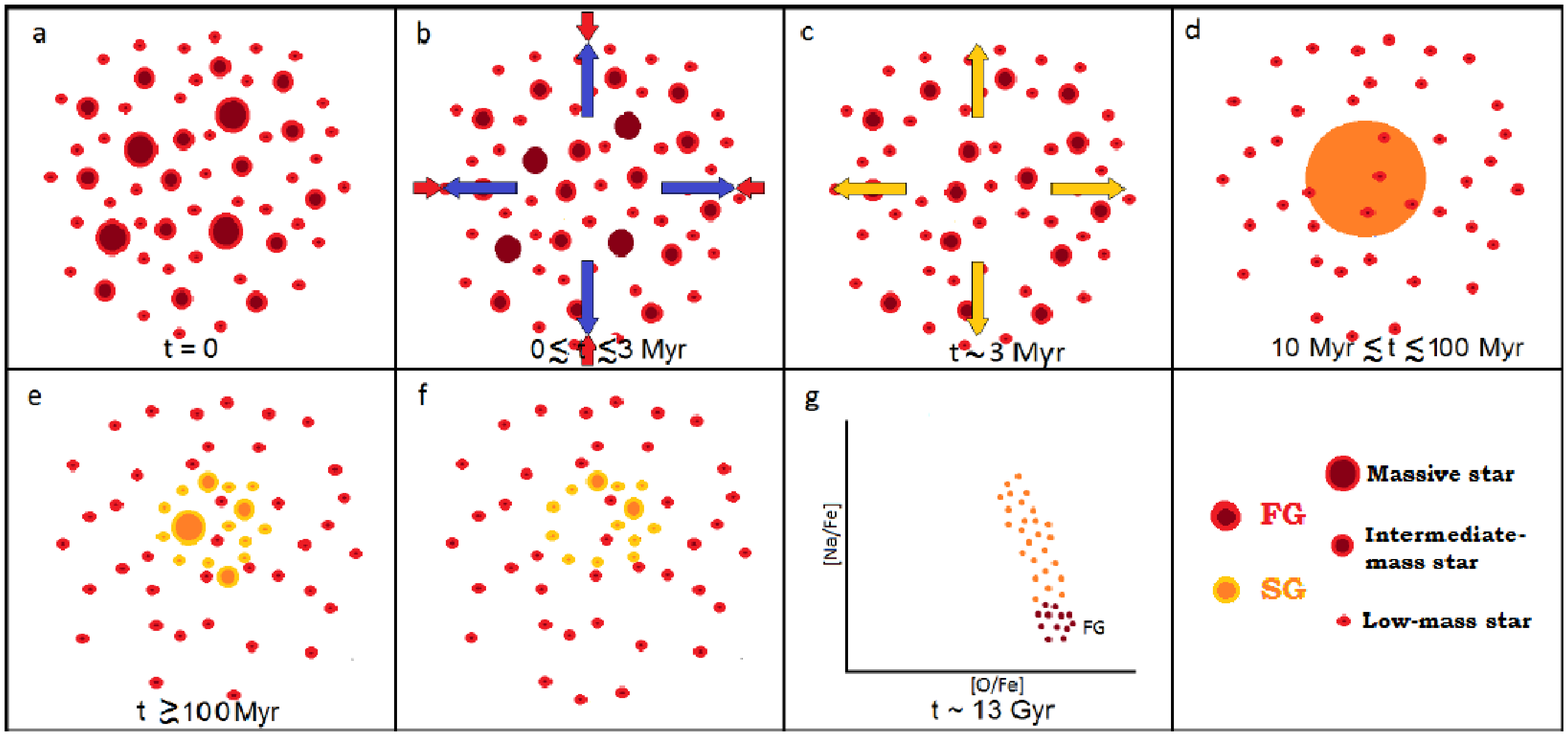}
      \caption{Schematic representation of our scenario for low-mass PSs. Stars of the first and second generations are represented as red and orange, respectively. Sizes indicate the different star masses, where large, medium, and small circles represent massive, intermediate-mass, and low-mass stars, respectively. Arrows show the gas flux, with the arrow color indicating the chemical composition following the same color scheme as for the stars of the different generations, and the arrow size being roughly proportional to the gas speed. Colored areas represent zones of gas accumulation. In each panel an approximate time for these events is given. Panels represent: a)~Formation of FG stars; b)~Pristine gas expulsion due to winds of FG massive stars; c)~Core-collapse SN ejecta are not retained in low-mass PSs; d)~Gas accumulation from winds of FG super-AGB and AGB stars; e)~Formation of SG stars; f)~SG core-collapse SNe period; g)~Present-day O-Na anticorrelation.}
      \label{FIGNonMassiveGC}
 \end{figure*}

\subsection{A Final Objection Against the Super-AGB Scenario for the Population Richest in He}

\citet{DErcole_etal2010} have shown that, in order to reproduce the observed O-Na anticorrelation, as well as the suggested He spread in NGC~2808, SG stars must form from super-AGB ejecta {\em only}, whereas in the case of M4 (NGC~6121) SG stars must be formed with a large amount of pristine gas mixed in. This is very surprising, since NGC~2808 is ten times more massive than M4 ($1.8\times 10^5 M_\odot $), whereas it should be easier for a massive PS than for a low-mass PS to retain pristine gas. In this case, M4 would have lost many more stars since its formation than did NGC~2808, implying a more massive PS in the past~-- but not so massive as to have retained any (unobserved) trace of metals from SN explosions. However, due to their deeper potential wells, massive PSs are expected to lose a smaller number of stars than less massive PSs. This is contrary to what would have been expected if M4 were related with a massive PS but NGC~2808 with a less massive one, unless the M4 progenitor has been affected by stronger dynamical interactions in the course of its lifetime.

Finally, we believe it is important to emphasize that, in our scenario, super-AGB and/or normal AGB stars {\em do} play an important role in the formation of multiple populations. In fact, as we shall see, while our scenario implies that these stars cannot be the progenitors of the most helium-rich population observed in NGC~2808 and $\omega$~Cen, it does also indicate that they are likely to be the progenitors of the stars with helium abundances falling in between the ``normal'' (i.e., non-He-enriched) and most He-rich stars.

 \begin{figure}[!t]
   \centering
      \includegraphics[width=9.0cm,height=9.0cm]{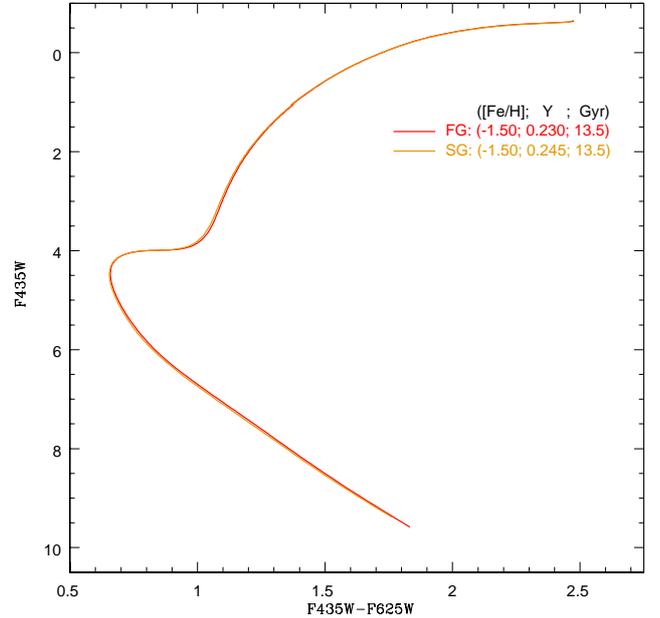}
      \caption{CMD of a low-mass PS after 13 Gyr. PGPUC isochrones represent the two surviving populations for this system: FG (red lines) and SG (orange lines).}
      \label{FIGNonMassiveGCIso}
 \end{figure}

 \begin{figure*}[t]
   \centering
      \includegraphics[width=18cm]{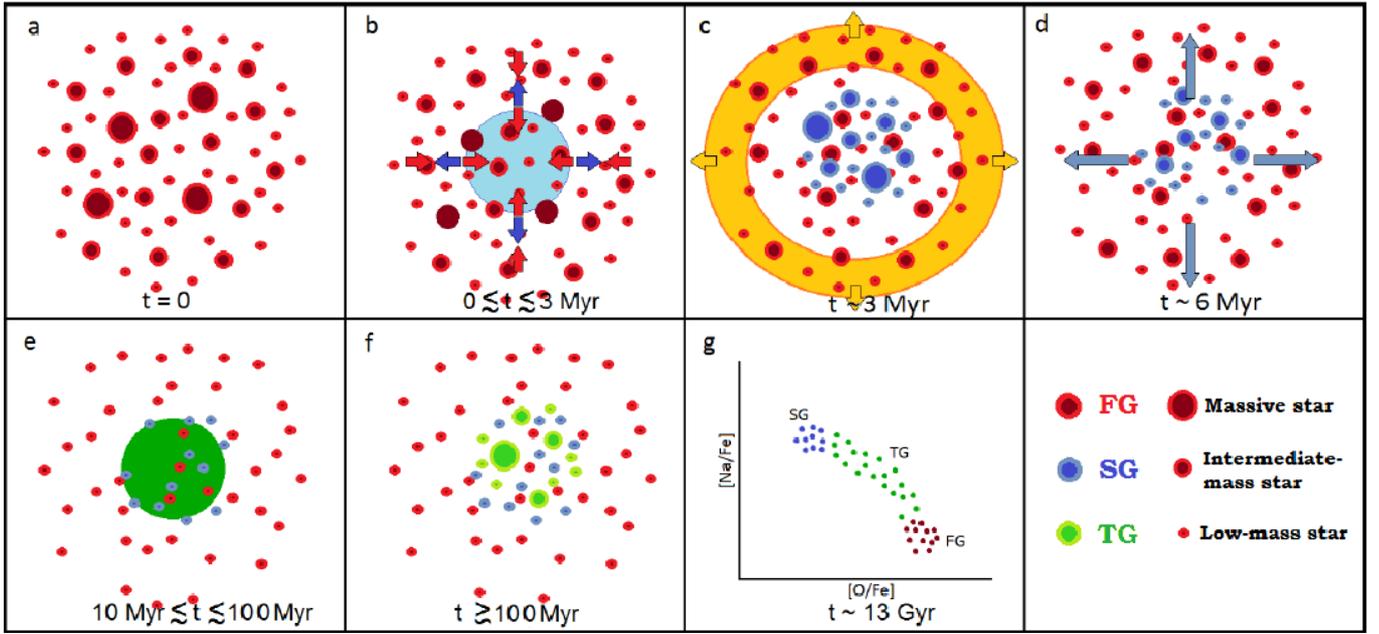}
      \caption{As in Fig.~\ref{FIGNonMassiveGC}, but for intermediate-mass PSs. Here the first, second, and third generation of stars are represented in red, blue, and green, respectively. Panels represent: a)~Formation of FG stars; b)~Gas accumulation from winds of FG massive stars and pristine gas; c)~First core-collapse SN explosions, which trigger the formation of SG stars and expel the gas that has not arrived to the core; d)~FG and SG SN explosions; e)~Gas accumulation from winds of super-AGB and AGB stars of the FG and SG; f)~Formation of TG stars;  g)~Present-day O-Na anticorrelation.}
      \label{FIGFairlyMassiveGC}
 \end{figure*}

\section{Our Scenario}
\label{ourmodel}

Our new scenario is divided in two stages, and is schematically described in Figs.~\ref{FIGNonMassiveGC}, \ref{FIGFairlyMassiveGC}, and \ref{FIGMassiveGC}). The first stage gives the basis for the formation of any GC, reflecting the conditions in the ISM that allow a gravitationally bound ``proto-GC'' to form (panels labeled ``a'' in Figs.~\ref{FIGNonMassiveGC}, \ref{FIGFairlyMassiveGC}, \ref{FIGMassiveGC}). In turn, the second stage depends on the PS's initial mass $M_I$, and can have three different outcomes. More specifically, i)~a low-mass PS only retains the mass ejected via slow winds from intermediate-mass stars; ii)~an intermediate-mass PS does not retain the core-collapse SNe ejecta, but does retain at least a fraction of the wind from massive stars; and iii)~a massive PS retains the winds of massive stars, and also the ejecta of core-collapse SNe.

\subsection{First Stage}

\begin{itemize}
\renewcommand{\labelitemi}{1.}
\item As usual, the first step for the formation of a GC begins with the gravitational collapse of a cloud with the {\em present-day} metallicity of the GC.  

\renewcommand{\labelitemi}{2.}
\item FG stars are then formed following a homogeneous distribution throughout the GC. They are initially embedded inside the ISM gas that was not used up to form stars (panels labeled ``a'' in Figs.~\ref{FIGNonMassiveGC}, \ref{FIGFairlyMassiveGC}, and \ref{FIGMassiveGC}). 

\renewcommand{\labelitemi}{3.}
\item Assuming a star formation efficiency ($\epsilon$) between 20\%, and 40\% \citep{Parmentier_etal2008}, the remaining gas has a mass between $0.8 \, M_I$ and $0.6 \, M_I$, but is distributed over a larger volume than is required for triggering a new local fragmentation process to form a new generation of stars. This implies a decrease in the local gas pressure for the same potential well, with the consequence that the remaining gas starts to fall again into the center of the cluster.

\renewcommand{\labelitemi}{4.}
\item The final step of this stage begins when massive stars eject their envelopes at high velocities, which collide with the falling gas. This leads to a decrease in the speed of the outbound massive star ejecta, and also to a decrease in the speed of the inbout gas (panels labeled ``b'' in Figs.~\ref{FIGNonMassiveGC}, \ref{FIGFairlyMassiveGC}, and \ref{FIGMassiveGC}). The chemical composition of the massive star ejecta is initially quite similar to the primordial one, however after a while (depending on the initial stellar mass) this ejecta consists of almost only helium \citep{Limongi_Chieffi2007}, without differences in heavy element abundances except mainly for those elements that participate in the CNO, NeNa, and MgAl cycles.

\end{itemize}

\subsection{Second Stage for Low-Mass PSs}
\label{SS_NMGC}

For low-mass PSs, the second stage takes place as follows (see also panels a through f in Fig.~\ref{FIGNonMassiveGC}):

\begin{itemize}
\renewcommand{\labelitemi}{5a.}
\item Low-mass PSs are unable to retain the FG massive star ejecta because of their shallower potential well, and the consequent low velocity of the infalling primordial gas. For the same reason, the PS center does not have enough material to transform all the kinetic energy of massive star ejecta into thermal energy. The formation of a viable star-forming cloud in the core is thus inhibited (panel b in Fig.~\ref{FIGNonMassiveGC}).

\renewcommand{\labelitemi}{6a.}
\item The next step is driven by the FG core-collapse SNe explosions, which clean completely the PS of the remaining primordial gas (panel c in Fig.~\ref{FIGNonMassiveGC}).

\renewcommand{\labelitemi}{7a.}
\item In this case, SG stars are formed only with (properly diluted) gas ejected by super-AGB and/or AGB stars (panels d and e), with the chemical composition of the ejecta depending in detail on the stellar mass. 

\renewcommand{\labelitemi}{8a.}
\item This is potentially a continuous process, with renewed cleansing of the intracluster gas taking place after each new star formation event (panel f in Fig.~\ref{FIGNonMassiveGC}). 
\end{itemize}

If the above scenario is correct, each successive stellar generation would leave a different signature in frequently observed diagnostic planes, such as the [Na/Fe] vs. [O/Fe] diagram. This diagram is schematically shown in panel g of Figure~\ref{FIGNonMassiveGC}. The anticipated impact on the HR diagram is shown in Figure~\ref{FIGNonMassiveGCIso}, where our new set of Princeton-Goddard-PUC (PGPUC) isochrones has been used \citep{Valcarce_etal2011}.

\subsection{Second Stage for Intermediate-Mass PSs (NGC~2808's Progenitor)}
\label{SS_FMGC}

A schematic representation of the second stage of the formation of intermediate-mass PSs, which are related to the progenitors of GCs like NGC~2808 (and possibly NGC~2419), is shown in panels b through f of Figure~\ref{FIGFairlyMassiveGC}, and can be summarized as follows:

\begin{itemize}
\renewcommand{\labelitemi}{5b.}

\item Due to the fact that intermediate-mass PSs have deeper gravitational potential wells than do low-mass PSs, the falling gas reaches a higher speed, and a fraction of the massive star ejecta is retained before their progenitors explode. However, in the outer part of the PS the massive star ejecta are trying to escape, while the pristine gas is infalling (panel b in Fig.~\ref{FIGFairlyMassiveGC}). Additionally, in the PS core, where the primordial gas has been accumulating, the gas ejected by massive stars is slowed down, transforming the kinetic energy into thermal energy, which delays further star formation.

 \begin{figure}[!]
   \centering
      \includegraphics[width=9.0cm,height=9.0cm]{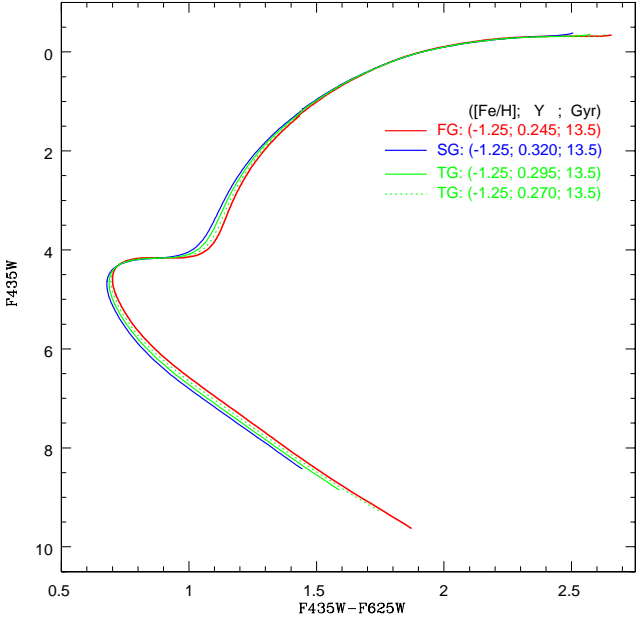}
      \caption{As in Figure~\ref{FIGNonMassiveGCIso}, but for an intermediate-mass PS. PGPUC isochrones represent the different populations presented for this PS: FG (red lines), SG (blue lines), and TG (green lines).}
      \label{FIGFairlyMassiveGCCMD}
 \end{figure}

\renewcommand{\labelitemi}{6b.}
\item When FG core-collapse SNe explode, their ejecta compress the center cloud, thus triggering the SG's star formation. These SG stars are highly enriched in helium due to the massive star ejecta, but at the same time they are not heavily enriched in metals. Assuming that SN explosions are nearly symmetrical and that their precursors are not too close to the center, only a small fraction of the SN ejecta, which is metal-enriched, will be mixed with the core cloud. This event also completely removes the outer mixed gas from the cluster (panel c in Fig.~\ref{FIGFairlyMassiveGC}), since the gravitational potential well is not deep enough and the mass of the outer falling gas is insufficient to retain the SNe ejecta. Thus, TG stars will accordingly {\em not} be metal-enriched.

\renewcommand{\labelitemi}{7b.}
\item Now, the ejecta of massive SG stars are not retained in the case of an intermediate-mass PS, because the infalling gas is only produced by intermediate-mass FG stars. If any such gas is initially retained, it will eventually be expelled by SG core-collapse SNe or FG type Ia SNe explosions, or will form but very few stars (panel d in Fig.~\ref{FIGFairlyMassiveGC}).

\renewcommand{\labelitemi}{8b.}
\item After this second cleansing of the cluster, a new cloud is starting to form in the cluster center, using the intermediate-mass stellar ejecta (mass lost at low velocity) from FG and SG stars (panel e in Fig.~\ref{FIGFairlyMassiveGC}). The chemical composition of this new cloud falls between both generations, as a consequence of the slope of the IMF (which favors low-mass stars) and the mass ratio between both generations. Here, the first stars of the third generation (TG) will be created.

\renewcommand{\labelitemi}{9b.}
\item Such star formation and cluster ISM cleansing stages continue, with each successive stellar population being chemically more similar to FG stars, while at the same time less numerous (panel f in Fig.~\ref{FIGFairlyMassiveGC}). 

\end{itemize}

In panel g of Figure~\ref{FIGFairlyMassiveGC} we show schematically the expected shape of the diagnostic [Na/Fe] vs. [O/Fe] diagram. The corresponding HR diagrams for this class of PS are shown in Figure~\ref{FIGFairlyMassiveGCCMD}.

\subsection{Second Stage for Massive PSs ($\omega$~Cen's Progenitor)}
\label{SS_MGC}

 \begin{figure*}
   \centering
      \includegraphics[width=18cm]{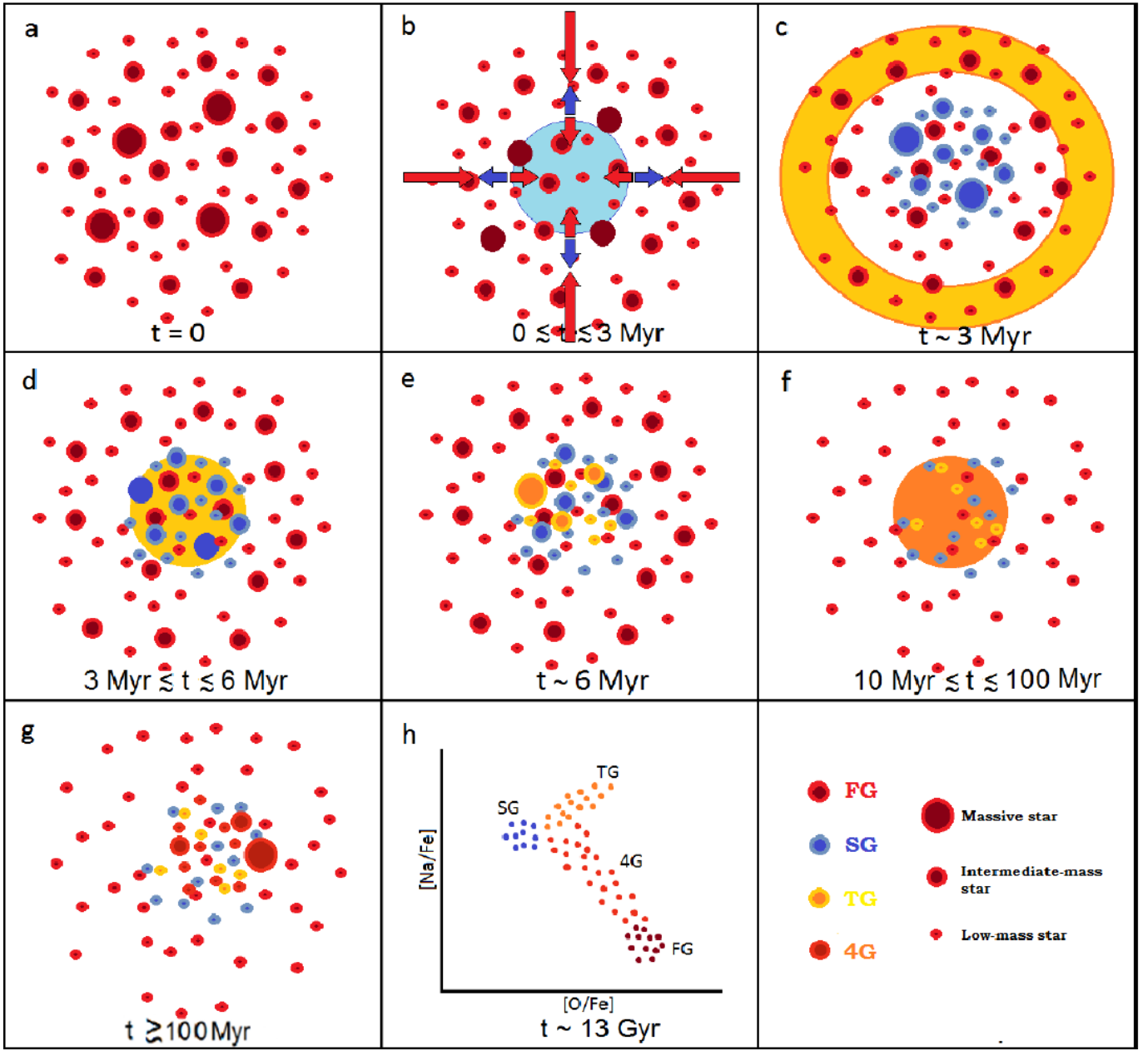}
      \caption{As in Figure~\ref{FIGNonMassiveGC}, but for massive PSs. Stars of the first, second, third, and fourth generations are represented as red, blue, yellow, and orange circles, respectively. Panels represent: a)~Formation of FG stars; b)~Gas accumulation, from a mix of winds of FG massive stars plus pristine gas; c)~First core-collapse SN explosions, which trigger the formation of SG stars; d)~Gas accumulation, from a mix of FG SN ejecta, winds of FG and SG massive stars, and gas remaining from the formation of SG stars; e)~Formation of TG stars, when SN explosions end; f)~Gas accumulation, from winds of super-AGB and AGB stars from all generations; g)~Formation of 4G stars; h)~{\em Present-day} O-Na anticorrelation.}
      \label{FIGMassiveGC}
 \end{figure*}

Figure~\ref{FIGMassiveGC} is a schematic representation of our scenario for massive PSs, which is explained in the following paragraphs. Here, one can identify $\omega$~Cen as a prototype of massive PSs~-- but other posssible examples include M22, M54, and Terzan~5 (see \S\ref{intro}). 

\begin{itemize}
\renewcommand{\labelitemi}{5c.}

\item As in $5b$, in the PS core a cloud has been formed that is highly enriched in helium. Its material comes from massive star ejecta and a fraction of the remaining gas that was not used to form FG stars. However, in the outer part of the GC the massive star ejecta is trying to escape, while the pristine gas is infalling with a velocity (and a total mass) that is higher than in the case of intermediate-mass PSs. 

\renewcommand{\labelitemi}{6c.}
\item The first core-collapse SNe explosions compress the core gas, which triggers the SG star formation episode (panel c in Fig.~\ref{FIGMassiveGC}). As in $6b$, these SG stars are highly enriched in helium (without however having significantly higher metal abundances). However, contrary to what happens in the case of intermediate-mass PSs, the higher potential well does allow the gas in the outer part of the cloud to be retained. Most of the SN ejecta tries to escape the cluster, merging with the falling gas in the process. This event efficiently mixes both gas components, and delays the moment of arrival of the further mixed gas to the core. 

\renewcommand{\labelitemi}{7c.}
\item After a while, the highly metal-enriched material is mixed with the gas which was not used to form SG stars, creating a new cloud in the core of the cluster (panel d in Fig.~\ref{FIGMassiveGC}). This cloud is also fed by SG massive stars and by both massive and intermediate-mass FG stars~-- and these provide the chemical ingredients that will characterize the cluster's third generation (TG) of stars (panel e in Fig.~\ref{FIGMassiveGC}). Unfortunately, this population has a helium abundance which is difficult to predict, because of the several sources of material that are involved. However, we expect that TG stars have some degree of helium enrichment.

\renewcommand{\labelitemi}{8c.}
\item Now, with three stellar generations in the cluster, the process of star formation continues, but each time with material mainly processed by less massive stars (super-AGB and/or AGB stars). As a result, the newborn stars belonging to the fourth generation (4G) will have a chemical composition which is a mix between the three preceding generations. However, in our scenario each new star that is formed will have a chemical composition more similar to that of the primordial gas, owing to two main factors: i)~The IMF slope, which implies more stars with low masses than massive stars, and ii)~The relatively small amount of mass used to form each star generation: indeed, in our scenario SG and TG stars are created using only a relatively small fraction of the total mass that was used to form FG stars. In other words, subsequent stellar generations will be more and more affected by the evolution of lower-mass stars, whose ejecta will not be chemically very different from that of the original PS gas (panels f and g in Fig.~\ref{FIGMassiveGC}).

\renewcommand{\labelitemi}{9c.}
\item Finally, a massive PS can lose the gas that has not been used to form stars in different ways, the most important mechanism probably being the interaction with larger structures~-- e.g., when a PS passes through the Milky Way. When this happens, at first only the intracluster gas is lost to the Galaxy.\footnote{According to \citet{Priestley_etal2011}, the mass evolution of the intracluster gas depends mainly on the PS mass, the PS velocity, the specific stellar mass loss rate, and the density of the medium where the PS is moving. However, to properly test our scenario, an extension of their study to larger PS masses, higher PS velocities, and denser underlying media would be needed.} Subsequently, however, due to the decrease of the PS's potential well, the sizes of the stellar orbits increase \citep[e.g.,][and references therein]{Moeckel_Bate2010,Decressin_etal2010,Trenti_etal2010}, and in this way the massive PSs can more easily lose their stars as well. If the gas expulsion is inefficient~-- e.g., when the massive PS has a lower interaction with larger structures~-- the PS does not lose a large number of stars because the gas has been retained, and thus ends up as a dwarf galaxy.

Of course, the strength of the interaction depends on the mass of the host galaxy. In the case of smaller galaxies (e.g., the Large and Small Magellanic Clouds), the possibility that the PS may lose mass due to interaction with the parent galaxy is much diminished, compared with the case of a galaxy like the Milky Way or M31. Conversely, the opposite may be true in galaxies that are more massive than the Milky Way, with the PSs potentially losing more mass in such massive galaxies \citep[see also ][]{Bekki2011}. Therefore, clusters with similar {\em present-day} masses in the Milky Way and the Magellanic Clouds are likely to be associated with more massive PSs in the former than in the latter galaxies \citep[but see][]{rcea10}. The end result is that the chemical inhomogeneities are expected to be less extreme in Magellanic Cloud clusters of a given (present-day) mass than in Milky Way clusters of similar (present-day) mass \citep[see also][]{Mucciarelli_etal2011b}~-- and vice-versa in the case of galaxies that are more massive than the Milky Way. This has important potential implications for the interpretation of the integrated light of GCs \citep[e.g.,][]{Kaviraj_etal2007, Rey_etal2007, Mieske_etal2008, Georgiev_etal2009, Coelho_etal2011, Peacock_etal2011}, potentially leading to parent galaxy mass-related effects.  

\end{itemize}

 \begin{figure}
   \centering
      \includegraphics[width=9.0cm,height=9.0cm]{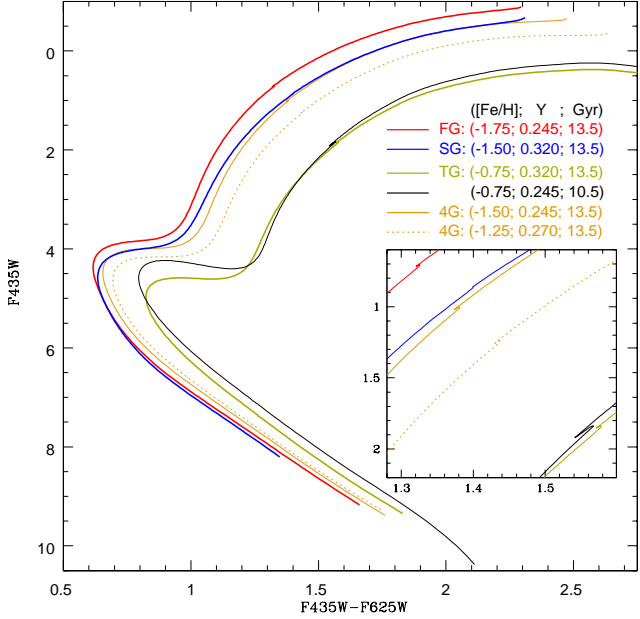}
      \caption{As in Fig. \ref{FIGNonMassiveGCIso}, but for a massive PS after 13 Gyr. PGPUC isochrones represent the different populations presented for this PS: FG (red lines), SG (blue lines), TG (gold lines), and 4G (orange lines). In black lines represents a possible younger TG population. Inner panel is a zoom-in at the RGB bump.}
      \label{FIGIsoMassiveGC}
 \end{figure}

Note that, according to our scenario, each subsequent stellar generation in $\omega$~Cen need {\em not} differ substantially in age. This is in contrast with early claims that $\omega$~Cen harbors stellar populations with an internal age spread that can reach several Gyr \citep[e.g.,][]{Hughes_Wallerstein2000, Rey_etal2004, Stanford_etal2006}. It should be noted, however, that these earlier results were obtained without taking into account the possibility of different levels of He enrichment in the different populations, which can lead to biases in the age estimates. In addition, it has also been recently suggested that the ages of the different populations in $\omega$~Cen are indeed very similar, perhaps to within 2~Gyr \citep{Sollima_etal2005b, Pancino_etal2011, DAntona_etal2011}. That a uniform-age solution is consistent with the cluster's CMD is clearly shown in Figure~8 of \citet{Sollima_etal2005b}, where a fit to high-quality ACS@HST photometry is shown in the constant-age/multiple-Y scenario.

As for the previous PS classes, in panel h of Figure~\ref{FIGMassiveGC} we show schematically the expected shape of the diagnostic [Na/Fe] vs. [O/Fe] diagram. The corresponding HR diagrams for this class of PS are shown in Figure~\ref{FIGIsoMassiveGC}.

In regard to the age issue and $\omega$~Cen's CMD, one can observe from Figure~\ref{FIGIsoMassiveGC} two important points: 

\begin{itemize}
\item The youngest SGB of the TG population with normal Y is not parallel to the SGB of an older FG population. Coeval SGBs, on the other hand, {\em are} parallel. The observations reveal nearly parallel SGBs for the different populations \citep[see, e.g., Fig 5. of][]{Villanova_etal2007}, thus supporting a small age spread in the cluster.   

\item The RGB bump of the TG population shows a greater spread in luminosity than the one of the FG population if these populations have the same helium abundances. However, if the TG is helium-rich, the spread in luminosities are predicted to be similar. That the observed spreads are quite similar is supported by Figure~10 of \citet{Rey_etal2004}.
\end{itemize}

As a final comment regarding the origin of the putative TG (i.e., the more metal-rich) component in $\omega$~Cen, we note that it clearly corresponds to a {\em discrete} component in this cluster. However, it is not clear to us how several Gyr can go by after the SN II explosions until such a metal-rich component is formed \citep[see also][]{rgea11}. SN~Ia do not appear to constitute a feasible solution, since the most metal-rich stars in $\omega$~Cen are also highly enriched in O and Na with respect to SG stars \citep{Johnson_Pilachowski2010,Marino_etal2011oCen,rgea11}. In fact, yields of SN~II \citep[see Fig.~8 in][]{Limongi_Chieffi2003} do show an increase in Na, but not in O~-- which means that, according to our scenario, TG stars will have [O/Fe] similar to SG stars, but higher [Na/Fe] and [Fe/H]. This is precisely the behavior that is observed in Figure~19 of \citet{Johnson_Pilachowski2010}. 

In summary, according to our scenario, in the case of a very massive progenitor, we expect that SG stars are He- and Na-rich, but O-poor (yields from a combination between pristine gas and massive stars before they explode), with a small enrichment in Fe (contribution from a small fraction of FG SN~II material). TG stars, on the other hand, are expected to be He-, Na-, and Fe-rich, but O-poor. Even though this model predicts separate distributions in the O-Na plane for the FG, SG and TG stars~-- a gap which becomes ``bridged'' when the 4G stars are formed (see Fig.~\ref{FIGMassiveGC}, panel h)~-- one should bear in mind that this is only a schematic representation, and the real chemical evolution can be more complex, especially due to the fact that 4G stars are formed with material processed by previous generations plus a fraction of pristine gas (if the latter has not been completely expelled). In fact, as is observed in the O-Na anticorrelation of $\omega$~Cen \citep[][]{Johnson_Pilachowski2010, Marino_etal2010}, there is an overlap between metal-poor and metal-intermediate branches in the O-Na plane. However, metal-rich stars (${\rm [Fe/H]}>-1.3$) are all Na-rich, which can also be associated to some degree of He enrichment.

\subsection{General Comments}

In this section, we have presented a scenario for the formation of multiple populations in GCs which makes several non-standard assumptions. The purpose of this subsection is to explain why these assumptions were made.

First, we incorporate into the cluster's evolving chemistry the ejecta of massive stars. Most frequently, however, the contribution of massive stars is ignored, due to the fact that their winds are very fast \citep[200 km/s $\le$ v $\le$ 2200 km/s;][]{Lamers_Nugis2002}. We decided to take these winds into account for four main reasons, namely:

\begin{itemize}
\item First, massive stars are clearly good candidates to explain populations with high helium abundances without an increase in metals, provided their ejecta is considered before they explode as core-collapse SNe. This is due to the fact that they go through evolutionary stages during which their atmospheres are comprised entirely of helium \citep[][and references therein]{Limongi_Chieffi2007, Pastorello_etal2008, Jeffery_Hamann2010}. 
Moreover, core-collapse SNe classification have divided them into three groups, depending on the elements observed in their spectras: i)~SNe Type II, whose spectra show H lines; ii)~SNe Type Ib, whose spectra shows He, but no H lines; and iii)~SNe Type Ic, whose spectra show neither H nor He lines. It has been proposed that these differences are due to the SNe progenitors' different masses, with the progenitors of SNe Type II being less massive than those of SNe Type Ib, which in turn are less massive than the SNe Type Ic progenitors \citep[see, e.g.,][for a recent review]{Leonard2010}. It appears accordingly that the progenitors of SNe Type Ib and Ic return to the ISM a large amount of helium before they explode.

\item In this sense, the material that arrives at the center of the PS is composed of a mix of pristine gas and the ejecta provided by the winds of massive stars. Before SNe explode, the yields of these winds are similar to the pristine composition early on, but almost completely comprised of He created via high-temperature H burning towards the end. Naturally, the PS cannot and does not select only the He-enriched wind ejecta to create SG stars; rather, SG stars are formed from a mix of pristine gas (i.e., gas that was not used to form FG stars, plus winds of FG stars) and He-rich winds.

\item If PSs have lost a large fraction of their mass in the course of their lives, and their present-day mass is of order 1\% of their initial mass only,\footnote{\citet{Decressin_etal2007b} have pointed out that, in order to reproduce the high fraction of SG stars, it is indeed necessary for the GC to have lost close to 96\% of all FG stars, unless a flat IMF for massive stars is adopted.} the escape velocity of the PS was ten times greater than the {\em present-day} escape velocity, thus making the retention of the massive stellar winds within the cluster's gravitational potential well much more feasible early on. Moreover, the presence of intracluster gas also plays an important role in decreasing the velocity of the ejected gas. For example, in the case of a GC~-- a group of stars without intracluster gas~-- the velocity of the ejected material decreases only by the action of the GC potential well, and so, if the initial velocity is greater than the escape velocity, that ejecta will be expeled from the GC. However, in the case of a PS~-- a group of stars {\em with} intracluster gas~-- the velocity of the ejecta of stars is decreased by the GC potential well, as well as due to the interaction with the intracluster gas \citep[e.g.,][]{Dopita1981, Dopita_Smith1986, Brown_Burkert_Truran1995, Kasliwal_etal2005}. This means that, even if the initial velocity of the ejecta is greater than the escape velocity, the ejected gas can still be retained by the PS. Naturally, this process would increase in importance with the mass of the PS, since more massive PSs have a larger amount of pristine gas that can interact with the ejecta. 

\item If pristine gas is required to form SG stars \citep[as implied by the Li and F abundances measured in GC stars; see, e.g.,][]{Prantzos_etal2007}, the latter must form before the first core-collapse SNe, otherwise the pristine gas has either been expelled or become metal-enriched. Massive star winds are accordingly the only viable source of gas enrichment. Even though one can think that SG stars will never be enriched in metals, it is conceivable that the amount of core-collapse SN explosions required to trigger the formation of these stars depends on the amount of gas in the center of the PS, which could implies an increase of the metallicity, as in the case of $\omega$~Cen \citep[e.g.,][]{Piotto_etal2005}.

\end{itemize}

It should also be emphasized that stars with extreme He abundance may naturally have extreme abundances of other elements, such as high Na and low O. Therefore, stars at the high-Na, low-O end of the O-Na anticorrelation (panel h in Fig.~\ref{FIGMassiveGC}, and panels g in Figs.~\ref{FIGFairlyMassiveGC} and \ref{FIGNonMassiveGC}) are those which are more naturally expected to have high He abundances. This is in the same sense as recently observed by \citet{Bragaglia_etal2010}. This would also be qualitatively consistent with the observed correlation between the presence and extent of these abundance anomalies and position along the horizontal branch \citep[e.g.,][]{Catelan_deFreitasPacheco1995, Gratton_etal2010, Marino_etal2011M4}. However, to more properly test this scenario, models of massive stars with mass loss for low metallicities and with time-dependent yields are required, but these are unfortunately not yet available in the literature. 

Regarding the ratio between FG stars and subsequent generations of stars, the following remarks are in order. If every new generation is formed in the core of the PS, then the number of FG stars decreases more than does the number of stars belonging to the subsequent generations, due to the preferential evaporation of outer stellar members \citep{Decressin_etal2008}. As \citet{Carretta_etal2009} have suggested, these expelled stars can then become a main component of the halo field. Moreover, the effect presented by \citet{Decressin_etal2008} can be intensified due to expansion of the PS driven by expulsion of the gas that was not used to form stars \citep{Moeckel_Bate2010, Decressin_etal2010, Trenti_etal2010}. Even though our schematic plots try to represent the effect of the expulsion of FG stars, these figures may to some extent also overestimate the number of FG stars.

Finally, an additional case of PS could in principle exist, bracketing our definitions of intermediate-mass and low-mass PSs, where only FG and SG stars, but not TG stars, are formed. Such a PS would be expected to be characterized by a large gap in the O-Na diagram, between O/Na-``normal'' and O-poor/Na-rich stars. To the best of our knowledge, no such cases have been found in the literature. Indeed, for such a scenario to materialize, a continuous mechanism responsible for removing the stellar ejecta of super-AGB and AGB stars should be present. This mechanism could be provided by SN Ia explosions; however, a fraction of this gas could still be retained if these explosions are not very close to the PS core, thus allowing the formation of TG stars with a mixed chemical composition.


\section{Quantitative Estimates}
\label{Estimations}

In this section we show a method to estimate the initial mass $M_I$ of the PS that is expected in our scenario, on the basis of the {\em present-day} observed ratio between the number of SG and FG stars ($R^{SG}_{FG}$). 

First, assuming a constant star formation efficiency $\epsilon$, independent of the chemical composition, the total mass used to form the FG is 

\begin{equation}
\label{eqMFG_MI}
M_{FG} = M_I \times \epsilon ,
\end{equation}

\noindent while the remaining pristine gas has a mass of  

\begin{equation}
\label{eqMg_MI}
M_g = M_I \times (1-\epsilon ).
\end{equation}

Then, the mass used to form the SG stars is 

\begin{equation}
\label{eqMSG}
M_{SG} = (M_g \times f_g + M_{ej-FG})\times \epsilon ,
\end{equation}

\noindent where $f_g$ and $M_{ej-FG}$ are the mass fraction of the pristine gas and the fraction of ejected mass of FG stars which has fallen to the core of the PS (and thus used up to form SG stars), respectively. We estimate the last term from 

\begin{equation}
\label{eqMejected}
M_{ej-FG} = M_{FG} \times I \times f_I,
\end{equation}

\noindent where $I$ is the (time-dependent) fraction of the total mass ejected by massive FG stars, and $f_I$ is the fraction of this ejecta that has arrived to the PS core~-- which we again assume to be entirely used up in the formation of SG stars. 

Now, the present-day (i.e., at time $t_{GC}$) number of FG stars is $N_{FG} \times \eta_{FG}$, where $N_{FG}$ is the initial number of FG stars and $\eta_{FG}$ is the fraction of the $N_{FG}$ stars that are still alive at time $t_{GC}$. This can be determined using the IMF ($\phi$) as follows: 

\begin{equation}
\label{eta}
\eta_x = \int_{0.1}^{x} \phi (m) \ dm \ \ \bigg/ \int_{0.1}^{120} \phi (m) \ dm,
\end{equation}

\noindent where $x$ is the highest mass value at time $t_{GC}$ for the relevant chemical composition. For the maximium stellar mass, we adopted a conservative limit for low metallicities \citep{Stothers_Chin1993}, while the lowest mass limit was chosen according to \citet{Kroupa_etal1993}.

On the other hand, the number of stars which are still bound to the GC after a time $t_{GC}$ is given by 

\begin{equation}
N_{FG-now} =\eta_{FG} \times (1- S_L) \times N_{FG},
\end{equation}

\noindent where $S_L$ is the fraction of FG stars that have been expelled from the PS by dynamical interactions with the Milky Way and other structures. Moreover, $N_{FG}$ is related with $M_{FG}$ as $N_{FG} = M_{FG} / \langle m\rangle_t$, where $\langle m\rangle_t$ is the mean stellar mass, which depends on the time and the chemical composition, except when $t_{GC}=0$ ($\langle m\rangle_0$). Accordingly,

\begin{equation}
N_{FG-now} =\frac{\eta_{FG} \times (1- S_L) \times M_{FG}} {\langle m\rangle_0}.
\end{equation}

For the SG, this number is 

\begin{equation}
N_{SG-now} = \frac{\eta_{SG} \times (1-S^{SG}_L)\times M_{SG}}{\langle m\rangle_0},
\end{equation}

\noindent where we have assumed that a fraction $S^{SG}_L$ of SG stars has also been lost. 

Finally, the ratio between $N_{SG-now}$ and $N_{FG-now}$ ($R^{SG}_{FG}$), which can be estimated from observations, is

\begin{equation}
\label{eq_ratio1}
  R^{SG}_{FG}  =  \frac{ \eta_{SG} \times \left[ (1-\epsilon ) \times f_g +  \epsilon \times I \times f_I \right]   }    { \eta_{FG} \times (1- S_L) }\times (1-S^{SG}_L),
\end{equation}

\noindent which interestingly does not depend directly on either $M_I$ or $M_{GC}$, but is related to these quantities indirectly through $S_L$ and $\eta$. In eq.~\ref{eq_ratio1} all variables (including, naturally, $R^{SG}_{FG}$) can be estimated (by models and/or by observations), except for $S_L$ and $S^{SG}_L$. Therefore, if a value for $S^{SG}_L$ is assumed,\footnote{Observations have recently shown that $\sim2.5\%$ of the halo stellar population is CN-strong, and thus can be associated to SG stars that once belonged to GCs \citep{Martell_Grebel2010}. Moreover, the N-body simulations of \citet{DErcole_etal2008} show that if SG stars are formed in the centers of GCs (corresponding to $\sim 15\%$ of the total population), after 13~Gyr the fractions of FG and SG stars that are lost are $\sim 99\%$ and $\sim 70\%$, respectively, with the end result that the GC population is dominated by SG stars ($\sim 80\%$ are SG stars).} $S_L$ can be derived as follows:

\begin{equation}
\label{eq_ratio2}
   S_L = 1 - \frac{ \eta_{SG} \times \left[ (1-\epsilon ) \times f_g +  \epsilon \times I \times f_I \right]   }    { \eta_{FG} \times R^{SG}_{FG} }\times (1-S^{SG}_L), 
\end{equation}

Now, in order to determine the $M_I$ of a PS once $S_L$ has been computed, we note the following: 

\begin{itemize}
\item $M_I$ is related to $M_{FG}$ and $\epsilon$ through eq.~\ref{eqMFG_MI}.

\item $M_{FG}$ is related to the {\em present-day} mass of FG stars ($M_{FG-now}$), $S_L$ and the {\em present-day} mass fraction of $M_{FG}$ ($\mathfrak{M}_{FG}$) by $M_{FG-now} = M_{FG} \times (1- S_L) \times \mathfrak{M}_{FG}$. The $\mathfrak{M}_{FG}$ value is determined considering only stars with masses lower than the maximum mass ($x$) of stars belonging to the FG that are still alive, or  

\begin{equation}
\mathfrak{M}_x = \int_{0.1}^{x} m \times \phi (m) \ dm \ \ \bigg/ \int_{0.1}^{120} m \times \phi (m) \ dm.
\end{equation}

\item $M_{FG-now}$ is related to $M_{GC}$ and the fraction of {\em present-day} FG stars as $M_{FG-now} = M_{GC} \times \left( N_{FG-now} / N_{now} \right)$, where $N_{now}$ is the total {\em present-day} number of stars in the GC, or equivalently, $M_{FG-now} = M_{GC} / \left( 1 + R^{SG}_{FG}\right)$. \footnote{This neglects the mass of the stellar remnants, which constitute $\sim30\%$ of the total GC mass after 12~Gyr, the exact remnant mass fraction depending on the detailed dynamical evolution history, metallicity, and the IMF \citep{Kruijssen_Lamers2008}. On the other hand, the mass-to-light ratio also depends on these same parameters, varying in the range $M/L_{V}\approx 1$ to $7\, M_{\odot} L_{\odot}^{-1}$ at an age of 12~Gyr \citep{Kruijssen_Lamers2008}. Since the uncertainty in $M/L_{V}$ is much higher than that brought about by neglecting the stellar remnants, we decided to avoid the latter in our first, and admittedly rough, PS mass estimates.} 

\end{itemize}

In summary, we find 

\begin{equation}
\label{eq_initialmass}
  M_I = \frac{M_{GC}}{\mathfrak{M}_{FG} \times \epsilon \times \left( 1 + R^{SG}_{FG}\right) \times (1- S_L)},
\end{equation}

\noindent which relates $M_I$ with $S_L$, which can be estimated from the observations using eq.~\ref{eq_ratio2} (assuming a value for $S^{SG}_L$).

\subsection{The Case of NGC~2808} 
\label{NGC2808}

 \begin{figure}
   \centering
      \includegraphics[width=9cm,height=9cm]{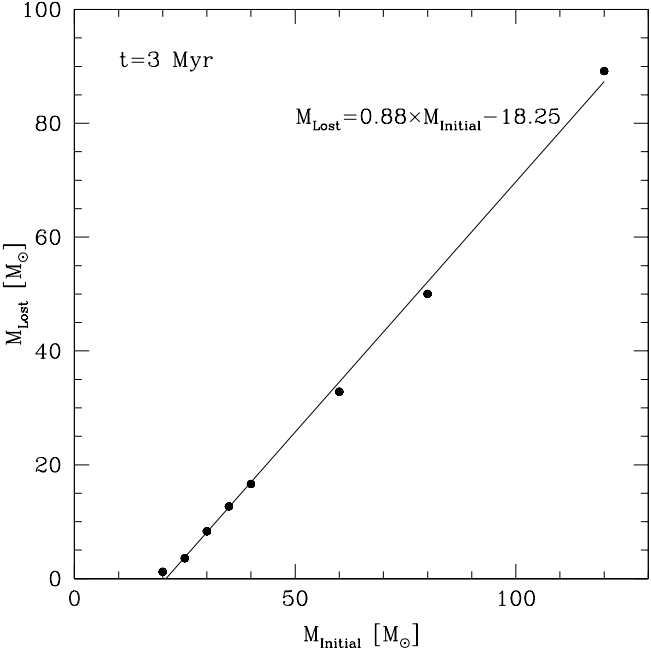}
      \caption{Relation between the initial mass and the total mass loss by massive stars after 3~Myr. Dots are from Table~2 of \citet{Limongi_Chieffi2006}, where a constant mass loss rate was assumed. The red line is our linear fit.}
      \label{FIGmasslost}
 \end{figure}

We now estimate $S_L$ and the $M_I$ of NGC~2808, a GC which we assume to be representative of intermediate-mass PSs.

We use a value of $\epsilon = 0.4 $ \citep{Parmentier_etal2008}, while the values for $\eta$ can be determined from eq.~\ref{eta} if the initial mass of the most massive star of each generation is known; the masses of low-mass stars are assumed not to decrease by a large amount. They are calculated knowing the chemical composition, and the age of each star generation. Here, we assume that both generations have the same age ($t_{GC}=12$~Gyr), metallicity ($Z=0.001$), and $\alpha$-element ratio ([$\alpha$/Fe] = 0.3), but different helium abundances \citep[$Y_{FG} = 0.245$ and $Y_{SG} = 0.345$;][]{Piotto_etal2007}. This implies maximum stellar masses of $\simeq 0.81$ $M_\odot$ and $\simeq 0.68$ $M_\odot$ for the FG and SG, respectively \citep[][]{Valcarce_etal2011}. Using a \citet{Kroupa_etal1993} IMF, we also find $\eta_{FG} = 0.50$ and $\eta_{SG} = 0.44$.

To determine $I$, we use the values from Table 2 of \citet{Limongi_Chieffi2007}. In this sense, as far as the present study is concerned, the most important conclusions are as follows:

\begin{itemize}

\item The first core-collapse SN (initial mass of $120 \, M_\odot$) explodes after 3~Myr.

\item Only stars more massive than $20 \, M_\odot$ lose a significant amount of mass in this short time interval. 

\item We obtain a linear relation between the initial mass of the stars and the mass lost after 3~Myr ($M_{Lost}$; see Fig.~\ref{FIGmasslost}), which reads as follows (with masses given in solar units): 

\begin{equation}
M_{Lost}({\rm 3 \, Myr}) = 0.88 \times M_{initial} - 18.25.  
\end{equation}

We assume, for simplicity, that massive stars lose their envelopes at a uniform rate during their lives. 

\end{itemize}

$I$ can thus be estimated as

\begin{equation}
I = \int_{20}^{120} M_{Lost}(m) \times \phi (m) \ dm \ \ \bigg/ \int_{0.1}^{120} m \times \phi (m) \ dm,  
\end{equation}

\noindent which at 3~Myr takes on a value of $0.015$. In other words, after 3~Myr stars more massive than $20\, M_\odot$ will have ejected 1.5\% of the total mass used to form the entire population of stars to which they are associated~-- low-mass stars included.

Finally, eq.~\ref{eq_ratio2} is reduced to

\begin{equation}
\label{RatioNGC2808}
  S_L  =  1 - \frac{ 5.28\times 10^{-1} \times f_g +  2.64\times 10^{-3} \times f_I } { R^{SG}_{FG} },
\end{equation}

\noindent which depends on $f_g$ and $f_I$, and where we have assumed, for simplicity, that a negligible number of SG stars has been lost. 

Due to the fact that our model predicts that SG stars should have low O abundances, to obtain an estimate of $f_g$ we use [O/Fe]$_{min}$ and [O/Fe]$_{max}$, which are the minimum and maximum abundances of this element relative to iron that are observed in the {\em present-day} GC. We proceed as follows:

\begin{itemize}
\item First, in a cloud of mass $M_{\rm u}$, the number of particles per unit mass of an element X ($N_{\rm X}$) is 

\begin{equation}
\label{eqNx}
N_{\rm X}= M_{\rm u} \times F_{\rm X},
\end{equation}

\noindent  where $F_{\rm X}$ is the mass fraction of element X.

\item Then, the ratio of the number of particles per unit mass between O and Fe in the progenitor cloud of SG stars is

\begin{equation}
\label{FracOFe}
\left( \frac{N_{\rm O}}{N_{\rm Fe}} \right)_{SG} = \frac{M_g \times f_g \times F_{\rm O,g} + M_{FG}\times I\times f_I \times F_{\rm O,FG}}{M_g \times f_g \times F_{\rm Fe,g} + M_{FG}\times I\times f_I \times F_{\rm Fe,FG}},
\end{equation}

\noindent where $M_g \times f_g$ and $F_{\rm X,g}$ are the mass used and the fraction by mass of the element X in the pristine gas, whereas $M_{FG}\times I\times f_I$ and $F_{\rm X,FG}$ are the mass used and the fraction by mass of the element X in the mass ejected by FG massive stars.

\item Now, if $F_{\rm Fe,g}=F_{\rm Fe,FG}=F_{\rm Fe}$, which means that massive stars do not eject material enriched in Fe in the first 3 Myr, eq. \ref{FracOFe} can be rewritten as

\begin{equation}
\label{FracOFe3}
\left( \frac{N_{\rm O}}{N_{\rm Fe}} \right)_{SG} = \frac{f_g \times (1-\epsilon) \times F_{\rm O,g} + \epsilon\times I\times f_I \times F_{\rm O,FG}}{\left[f_g \times (1-\epsilon)+ \epsilon \times I \times f_I\right]\times \epsilon\times F_{\rm Fe}},
\end{equation}

\noindent where $M_g$ and $M_{FG}$ are related to $M_I$ using eqs.~\ref{eqMFG_MI} and \ref{eqMg_MI}. Relating $F_{\rm O}$ and $F_{\rm Fe}$ to $N_{\rm O}$ and $N_{\rm Fe}$, respectively (eq.~\ref{eqNx}), eq.~\ref{FracOFe3} is equivalent to

\begin{equation}
\label{FracOFe4}
\left( \frac{N_{\rm O}}{N_{\rm Fe}} \right)_{SG} = \frac{1}{\epsilon \times (1+c)} \times \left[ \left(\frac{N_{\rm O}}{N_{\rm Fe}}\right)_g + c  \left(\frac{N_{\rm O}}{N_{\rm Fe}}\right)_{FG} \right],
\end{equation}

\noindent where
\begin{equation}
c = \frac{\epsilon\times I \times f_I}{(1-\epsilon)\times f_g}.
\end{equation}

\item Finally, eq.~\ref{FracOFe4} can be transformed to relative abundances,

\begin{equation}
10^{{\rm [O/Fe]}_{SG}} = \frac{1}{\epsilon \times (1+c)} \times \left[10^{{\rm [O/Fe]}_{g}} + c\times 10^{{\rm [O/Fe]}_{FG}} \right],
\end{equation}

\indent which means that the ejected gas by FG massive stars must have an O abundance of

\begin{equation}
{\rm [O/Fe]}_{FG} = \log \left[ \frac{\epsilon \times (1+c)\times10^{{\rm [O/Fe]}_{SG}}-10^{{\rm [O/Fe]}_{g}}}{c} \right].
\end{equation}

Note that the fraction $f_I/f_g$ that is used to compute $c$ is constrained by 
\begin{equation}
\label{eqConstrain}
\frac{f_I}{f_g} \ge \frac{1-\epsilon}{\epsilon \times I} \times \left( \frac{10^{\,\Delta{\rm [O/Fe]}}} {\epsilon} - 1 \right),
\end{equation}

\noindent where $\Delta{\rm [O/Fe]}={\rm [O/Fe]}_{g}-{\rm [O/Fe]}_{SG} $. Here we have obtained a relation between the fraction $f_g$ of primordial gas used to form SG stars and the fraction $f_I$ of gas ejected by FG stars that are used to form SG stars. Note that, by definition, neither of these values can exceed unity. 
\end{itemize}

 \begin{figure}
   \centering
      \includegraphics[width=9cm,height=9cm]{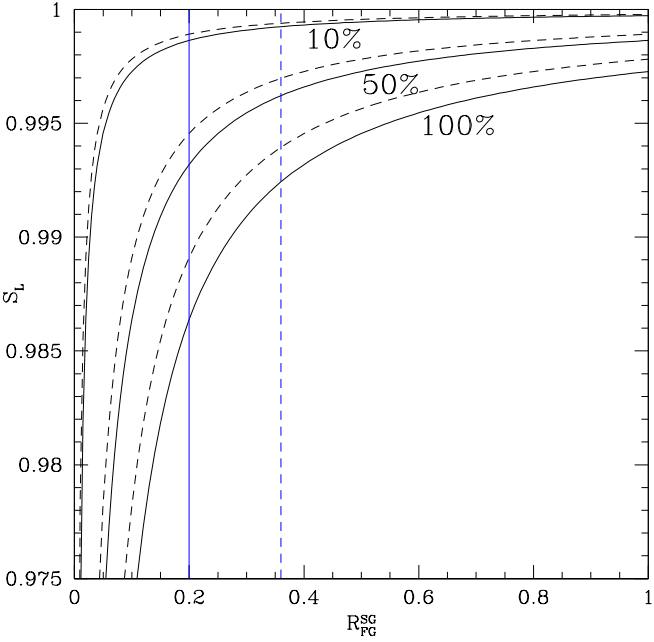}
      \caption{Fraction of FG stars that an intermediate-mass PS must lose ($S_L$) to reproduce the {\em present-day} ratio between the number of SG and FG stars. Black curves represent different mass fractions of gas ejected by massive FG stars in 3~Myr (values of $f_I$ in percentages) used to form the SG stars when $\epsilon$ is 40\% and when $S^{SG}_L$ is negligible (solid lines) or 20\% (dashed lines). The blue vertical lines represent the observed ratio between SG and FG stars ($R^{SG}_{FG}$) estimated from the number of red and blue MS stars \citep[solid line, from][]{Piotto_etal2007} and from the O-Na anticorrelation \citep[dashed line, from][]{Carretta_etal2009}, as observed in the case of NGC~2808.}
      \label{FIGequationNGC2808}
 \end{figure}

One can estimate the ratio in eq.~\ref{eqConstrain} by assuming that [O/Fe]$_g$ and [O/Fe]$_{SG}$ are directly given by [O/Fe]$_{max}$ and [O/Fe]$_{min}$, respectively, as indeed implied by our model. Using the abundance ratios derived by \citet{Carretta_etal2006} in the case of NGC~2808 ([O/Fe]$_{max}=0.4$ and [O/Fe]$_{min}=-1.0$), we obtain $f_I\ge 6180\times f_g$, where, as in eq. \ref{RatioNGC2808}, we used $I=0.015$ and $\epsilon=0.4$. This implies that, if SG stars in NGC~2808 were formed using 100\% of the ejected mass by massive FG stars (corresponding to 1.5\% of $M_I$), the maximum mass fraction of the primordial gas that can be used to form SG stars must be lower than 0.02\%, or else the observed values of [O/Fe] cannot be reproduced.\footnote{Observationally, one finds that the abundances of such fragile elements as Li and F are correlated with O \citep[and anticorrelated with Na; e.g.,][]{Pasquini_etal2005,Smith_etal2005, Denissenkov_etal2006, Shen_etal2010}. However, \citeauthor{Shen_etal2010} conclude that the slope of the observed Li-O correlation is not what one would expect from simple pollution scenarios, claiming that the polluting gas must be somewhat enriched in Li. If so, and as pointed out by \citeauthor{Shen_etal2010}, this would rule out massive stars as the main polluters, unless mechanisms can be found through which these massive stars can produce Li. Contrary to their conclusions, however, the chemical evolution models by \citet{klea11} account very well for the observed Li abundances in NGC~6397, using massive stars between $20$ and $120\, M_{\odot}$ as the polluters~-- provided their ejecta are suitably diluted with pristine material. Clearly, the chemical evolution of Li is far from being straightforward, as the remarkably constant Li abundance with metallicity observed by \citet{Monaco_etal2010} in $\omega$~Cen also shows. In addition, it is important to bear in mind that it is not straightforward to interpret Li abundance observations, given the many ill-constrained physical mechanisms that may play a relevant role in defining surface Li abundances in present-day stars \citep[e.g.,][]{Mucciarelli_etal2011}.
} 

Even though the derived limit for $f_g$ implies that O is completely depleted in the gas ejected by massive FG stars,\footnote{In fact, massive stars do produce oxygen, albeit to a lower extent than they produce iron. However, the material chemically enriched in elements heavier than C, He, and H, in the case of the progenitors of SNe Type Ic, Type Ib, and Type II, respectively, is not expelled via winds, but rather as SNe ejecta. This means that only PSs that can retain the ejecta of SNe can also have an enrichment in such heavier elements. This is in fact  observed in the O-Na anticorrelation of $\omega$~Cen, when abundance data are separated into bins of metallicity \citep[][]{Johnson_Pilachowski2010, Marino_etal2011oCen}.} it can be used to constrain the minimum value of $S_L$ depending on $f_g$ (or $f_I$). Then, using eq.~\ref{RatioNGC2808}, the fraction of FG stars that have been expelled is

\begin{equation}
  S_L \ge 1 - 2.73\times10^{-3}\times f_I \,/\, R_{FG}^{SG}.
\end{equation}

\noindent Figure~\ref{FIGequationNGC2808} shows the minimum fraction of FG stars that NGC~2808 has lost, according to our model. In the specific case of this GC, $R^{SG}_{FG}$ can be determined by two methods:

\begin{itemize}
\item Using the MS split \citep{Piotto_etal2007}, where 63\% of the population is in the reddest MS, while 13\% is in the bluest MS, and both populations can be associated to FG and SG stars, respectively~-- thus implying an $R^{SG}_{FG}{\rm (MS)}\approx0.20$. However, this method is the minimum value for $R^{SG}_{FG}$ because, according to our scenario, an unknown fraction of the stars in the redder MS were formed after some tens of Myr from material processed by super-AGB or AGB stars (thus belonging to later stellar generations; see Fig.~\ref{FIGFairlyMassiveGC}). These young stars in the redder MS must have a chemical composition (and in particular a helium abundance) similar to FG stars, but still showing abundance variations in some of the light elements.

\item Using the O-Na anticorrelation, where stars with low Na abundances belong to the FG and stars with low O abundances belong to the SG, whereas stars that belong to neither of these groups are TG stars. Following \citet{Carretta_etal2009} definitions for primordial, intermediate, and extreme populations, $R^{SG}_{FG}$ can be determined dividing the fractions of ``extreme'' stars by the fraction of ``primordial'' stars. This gives an $R^{SG}_{FG}{\rm (O-Na)}=0.36$.
\end{itemize}

In Figure~\ref{FIGequationNGC2808}, these $R^{SG}_{FG}$ ratios are represented by blue vertical lines, with the solid line corresponding to the MS CMD constraint, and the dashed one to the O-Na diagram constraint. If SG stars were formed using the total mass of massive FG stars ($f_I=100\%$), the minimum $S_L\sim 0.992$ for $R^{SG}_{FG}{\rm (O-Na)}$, while for $R^{SG}_{FG}{\rm (MS)}$ the minimum $S_L\sim 0.986$. However, if $f_I=10\%$, the minimum $S_L$ values are $\sim 0.999$ and $\sim 0.998$ for $R^{SG}_{FG}{\rm (O-Na)}$ and $R^{SG}_{FG}{\rm (MS)}$, respectively. Even though these values are high, the initial-to-final mass ratio of GCs depends on the orbit of each PS around the Milky Way, which can completely destroy them if the interaction is strong enough \citep[e.g.,][]{Kruijssen_etal2011}. 

Using the range of $S_L$ values determined with $f_I=1$ and eq.~\ref{eq_initialmass}, the minimum $M_I$ of NGC~2808 is estimated between $3.8$ and $5.9 \times 10^8 M_\odot$, where we used $\mathfrak{M}_{0.81} =0.504$.\footnote{This assumes that SG stars have not been lost during the PS's evolution. If, on the other hand, 70\% of SG stars have been expelled from the PS, as suggested by \citet{DErcole_etal2008}, NGC~2808's minimum $M_I$ values are increased, leading to masses between $1.3$ and $2.0\times 10^9 M_\odot$.} In fact, the large fraction of cluster mass lost makes our scenario more plausible, because the escape velocity of the PS of NGC~2808 was between $\sim860$ and $\sim1060$~km/s (and possibly higher, depending on how many SG stars are lost), instead of the {\em present-day} $\sim 50$~km/s~ \citep{Gnedin_etal2002}. Unfortunately, there are no estimates of the $M_I$ for NGC~2808 in the literature. However, the orbit of this GC suggests a much more massive PS than the present-day GC mass, due to interaction with the Milky Way \citep{Casetti-Dinescu_etal2007}. 

Another important point that should be emphasized is that our scenario naturally explains the discrete nature of the stellar populations observed in NGC~2808. This is due to the fact that, after the first core-collapse SN explosion and the triggered formation of SG stars, there will be a period of time without star formation, due to subsequent FG core-collapse SNe, which are interspersed with SG core-collapse SNe, followed by SNe Type Ia. In fact, one should expect that this process is repeated every time that a new generation is born, but every time with a chemical composition more similar to the FG. Accordingly, the observational detection of the differences between subsequent populations (TG in this case) becomes increasingly more difficult. 

\subsection{The Case of $\omega$~Cen}

 \begin{figure}
   \centering
      \includegraphics[width=9cm,height=9cm]{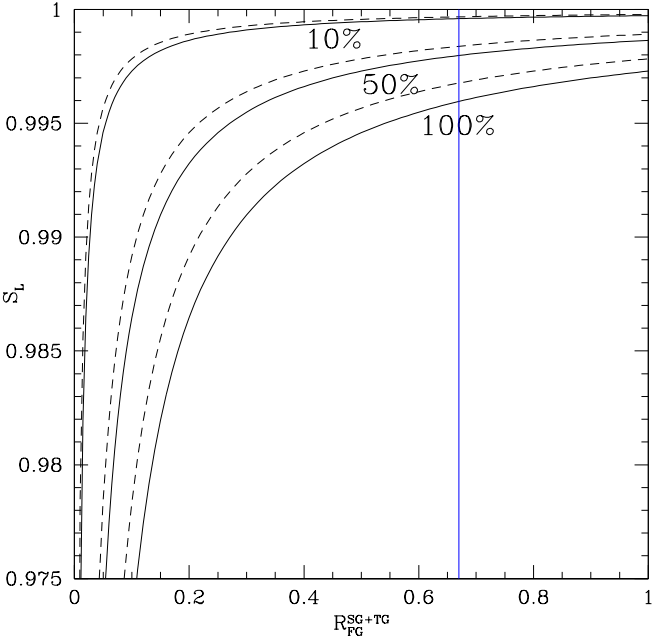}
      \caption{As in Figure~\ref{FIGequationNGC2808}, but in the case of $\omega$~Cen. The blue vertical line represents $R^{SG+TG}_{FG}=0.67$ (see text).}
      \label{WCenFIGequation2}
 \end{figure}

As usual, the case of $\omega$~Cen is more complex than for other GCs, due to the fact that in its CMD and in spectroscopic analyses alike a large numbers of populations is observed (\S1). However, we used the ratios of the different populations given in Table~4 of \citet[][]{Villanova_etal2007} to estimate the initial mass of this GC, as required to form SG and TG stars from gas previously processed by FG massive stars (plus a fraction of pristine gas).

In the formation scenario for massive PSs, which is the case assumed for $\omega$~Cen, there is only a small 
difference in metallicity between FG and SG stars, but a sizeable difference in the helium abundance. As in the case of NGC~2808, we associate a large fraction of the reddest MS of the cluster to FG stars (57\%), while SG stars are associated to the bluest MS (33\%). However, in this scenario the gas ejected by massive stars that is not used to form SG stars is used to form TG stars~-- which, according to our scenario, {\em will} be more metal-enhanced than the SG. We associate these putative TG stars to the MS-a component (containing 5\% of all MS stars) of \citet[][]{Villanova_etal2007}, which is suggested to be related to the cluster's most metal-rich RGB sequence \cite[RGB-a in][]{Sollima_etal2005}. With this in mind, we rewrite eq.~\ref{eq_ratio1} as

\begin{eqnarray}
\label{eq_ratioWcen}
  R^{SG+TG}_{FG} & = & \frac{N_{SG-now}+N_{TG-now}}{N_{FG}}. 
\end{eqnarray}

In section \ref{NGC2808}, we estimated a value for $I = 0.015$ which could have been overestimated. In the case of $\omega$~Cen, we again use the same value, though in this case it could instead represent an underestimate, due to the fact that after 3~Myr massive FG stars continue to return gas to the medium that is used to form SG and TG stars. We also use the same value of $\eta_{FG}$ estimated for NGC~2808, and assume that $\eta_{SG+TG} \approx \eta_{SG}$ ($m_{max-SG} \approx 0.68 \, M_\odot$), due to the small percentage of TG stars. Finally, using eq.~\ref{eqConstrain}, we obtain $f_I\ge 7800 \, f_g$, which determines a minimum value of $S_L$ for $\omega$~Cen of

\begin{equation}
  S_L \ge 1 - 2.71\times10^{-3}\times f_I \,/\, R_{FG}^{SG+TG}.
\end{equation}

In Figure~\ref{WCenFIGequation2} is shown the relation between $S_L$ and $R^{SG+TG}_{FG}$, where the blue vertical line corresponds to $R^{SG+TG}_{FG}=0.67$, as obtained from the \citet[][]{Villanova_etal2007} number counts. In the case where $f_I=1$, the fraction of FG stars expeled must be higher than 99.6\% of the initial number, implying a $M_I \approx 2.2\times 10^9 M_\odot$~-- which is between $\approx 4$ and 17 times higher than the estimates obtained on the basis of N-body simulations \citep[e.g.,][with favored $M_I$ values of $1.3$, $5.8$, and $5.0\times 10^8 M_\odot$, respectively]{Bekki_Freeman2003, Mizutani_etal2003,Wylie_deBoer_Cottrell2009}.\footnote{Here too we have assumed that SG stars have not been lost during the PS's evolution. If 70\% of SG stars have been expelled from the PS \citep{DErcole_etal2008}, $\omega$~Cen's $M_I$ value is increased to $7.4\times 10^9 M_\odot$, leading to overestimates in $M_I$ by factors in the range between $\approx 13$ and 57 with respect to the aforementioned N-body studies.} While our estimates of $M_I$ thus appear higher than the most likely (dynamical) value for this PS, we must stress that more detailed computations, incorporating the ingredients described in the next subsection, are clearly required before we are in a position to provide more than only a rough estimate of a PS's $M_I$ value.

\subsection{Required Improvements}

In previous sections we have related two GCs to a massive PS ($\omega$~Cen) and to an intermediate-mass PS (NGC~2808). However, we have not related any GC to low-mass PSs, due to the difficulty pointed out in \S\ref{DErcoleModel}, namely the treatment of Type Ia SNe. To solve this problem, models of GC evolution are needed which must take properly into account: i)~SNe explosions taking place at different positions inside the GCs; ii)~Interactions between gases; and iii)~The fraction of binary systems which are possible SN Type Ia progenitors.

More in general, a more precise determination of the following parameters is also required:

\begin{itemize}
\item $\epsilon$: the observed values of this parameter vary between 0.2 and 0.4 \citep{Parmentier_etal2008}, which affects estimates of $M_I$, higher values implying more mass ejected by massive FG stars, which can accordingly be used to form more SG stars.

\item IMF: As shown by (e.g.) \citet{Skillman2008}, for very low metallicities the IMF predicts more massive stars. However, a proper evaluation of IMFs for metallicities similar to those observed in GCs is required \citep[e.g.,][]{Sollima_etal2007}, including the case of high-helium environments. 

\item $I$: better estimates of this parameter require evolutionary tracks for massive stars with mass loss for low metallicities, along with their yields \citep[e.g.,][for intermediate-mass stars]{Herwig2004}. Moreover, due to the fact that such winds are strongly dependent on the metallicity (low-metallicity stars releasing a lower amount of mass), it is possible that a trend between subsequent generations and the metallicity exists. However, it is not possible to estimate if this trend is more or less sharp with the metallicity, since metallicity also changes the other two important parameters used in our approach: the IMF and $\epsilon$ \citep[e.g.,][]{Skillman2008,sdea11}.  

\item Other ingredients: there are many other possible mechanims that may play an important role in the formation of multiple populations with high helium abundances, but which we have not considered in our analysis. This includes, e.g., massive binary stars \citep{De_Mink_etal2009}, stellar collisions \citep{Sills_Glebbeek2010}, and ``localized enrichment'' \citep{Marcolini_etal2009}, among others (see also \citeauthor{gp04} \citeyear{gp04}; \citeauthor{gs06} 2006, \citeyear{gs10}; and \citeauthor{rdg10} \citeyear{rdg10} and \citeauthor{sm11} \citeyear{sm11} for recent reviews). Further work incorporating these potentially important ingredients is strongly encouraged.  
\end{itemize}

Finally, a solid model for the evolution of GCs and their multiple populations must consider the evolution of the PS structure after the expulsion of pristine gas, which leads to an expansion of the PS and a period of subsequent star loss \citep{Moeckel_Bate2010}. Moreover, in a model where massive stars are considered to play an important role in the evolution of GCs, the formation of low-mass stars cannot be assumed as instantaneous, due to the fact that proto-low-mass stars have contraction times of several tens of Myr \citep{Bernascon_Maeder1996}, and thus could become contaminated in that period of time by the winds of massive stars \citep{Newsham_Terndrup2007, Tsujimoto_etal2007}. 

\section{Summary}
\label{summary}

In this paper, we have summarized some of the most recent scenarios which try to explain the formation of multiple populations in GCs. We have also presented a new scenario, whose chief difference with respect to previous scenarios is the classification of GCs by their {\em initial} mass: i)~The most massive PSs can retain the ejecta of massive stars (winds and core-collapse SN ejecta); ii)~Intermediate-mass PSs can only retain the winds of massive stars, but not the SNe ejecta; and iii)~Low-mass PSs can only retain the slow winds of intermediate-mass stars \citep[see also][]{Bekki2011}. 

To avoid the classic argument against the possibility of using the ejecta of massive stars because of the high velocity of their winds (from hundreds to some thousands of km/s), in our scenario we postulate that: i)~GCs were (much) more massive in the past, with escape velocities of some hundreds of km/s, a large percentage of FG stars and the pristine gas having been expelled by core-collapse SNe and/or SNe Type Ia, and the interaction with their host galaxy; and ii)~After the formation of the FG of stars, the remaining pristine gas begins to fall to the center of the PS, interacting with massive stellar winds that try to escape the PS. An important aspect of our scenario is that it can naturally explain the discreet nature of FG and (especially) SG stellar populations, as have been observed in GCs. 

In general, it is exceedingly difficult to estimate the PS mass $M_I$ directly from the GC's present-day mass $M_{GC}$, because $M_{GC}$ depends on the GC's past dynamical history, including interactions with the Galaxy. On the other hand, our scenario allows order-of-magnitude estimates from different empirical estimates of the present-day number ratio between {\em present-day} FG and subsequent stellar generations that may be present in individual GCs. This leads to the minimum $M_I$ values for NGC~2808 in the range between about $3.8$ and $5.9 \times 10^8 \, M_\odot$, whereas for $\omega$~Cen we obtain $2.2\times 10^9 \, M_\odot$. In the case of NGC~2808, no previous $M_I$ estimates are available, but its orbit does suggest a massive progenitor \citep{Casetti-Dinescu_etal2007}. In the case of $\omega$~Cen, the mass implied by our scenario is higher than obtained on the basis of N-body simulations \citep{Bekki_Freeman2003,Mizutani_etal2003,Wylie_deBoer_Cottrell2009}. Further work is clearly required, before we are in a position to provide more than a rough estimate of a PS's $M_I$ value, based on present-day observables. 

Our scenario can also explain why the O-Na anticorrelation does not follow the same pattern for all GCs, where extreme high-Na/low-O stars are SG stars formed with material processed by massive stars and a fraction of the pristine gas (massive and intermediate-mass PSs), while normal high-Na/low-O are SG formed with material ejected by super-AGB or AGB stars (low-mass PSs). Subsequent generations are formed with a mixture of material from pre-existing generations, where the last stars to form (i.e., the youngest stars) are more chemically similar to FG stars.

Naturally, we are well aware that our scenario represents but a first ``toy model'' that will require extensive improvements, including three-dimensional hydrodynamical modeling, before it is in a position to provide more detailed, quantitative predictions regarding the nature and properties of the multiple populations that are now commonly observed in GCs. In this sense, detailed numerical investigations, such as the one recently carried out by \citet{Bekki2011}, are strongly encouraged.

\subsection*{Acknowledgments}
We warmly thank Allen V. Sweigart for many helpful comments and discussions throughout the course of this project. Useful discussions with Alvio Renzini, Santi Cassisi, Alejandro Clocchiatti, Andreas Reisenegger, and Manuela Zoccali are gratefully acknowledged. We are also grateful to the anonymous referee, whose comments and suggestions have helped improve the presentation of our results. Support for A.A.R.V. and M.C. is provided by the Ministry for the Economy, Development, and Tourism's Programa Inicativa Cient\'{i}fica Milenio through grant P07-021-F, awarded to The Milky Way Millennium Nucleus; by Proyecto Basal PFB-06/2007; by FONDAP Centro de Astrof\'{i}sica 15010003; and by Proyecto FONDECYT Regular \#1110326. A.A.R.V. acknowledges additional support from Proyecto ALMA-Conicyt 31090002, MECESUP2, and SOCHIAS.

\bibliographystyle{aa}
\bibliography{aavalcarcebib}

\end{document}